\documentclass[doublecol,linenumbers]{epl2}
\usepackage{amsmath}
\usepackage{amssymb}
\usepackage{amsfonts}
\usepackage{graphicx}
\usepackage{dcolumn}
\usepackage{bm}
\usepackage{sidecap}
\usepackage{float}
\usepackage{parskip}
\usepackage{grffile}

\usepackage{color}
\usepackage{hyperref}
\usepackage[normalem]{ulem}
\usepackage[table]{xcolor}
\usepackage{booktabs}
\usepackage[capitalize]{cleveref}
\definecolor{mygreen2}{RGB}{0, 126, 0}

\graphicspath{{./}{images/}}

\title{Wheel graph strategy for PEV localization of networks}

\author{Sarika Jalan\inst{1} and Priodyuti Pradhan\inst{1}}
\shortauthor{S. Jalan \etal}

\institute{
  \inst{1} Complex Systems Lab, Discipline of Physics, Indian Institute of Technology Indore, Khandwa Road, Simrol, 453552, Indore, India
}
\abstract{Investigation of eigenvector localization properties of complex networks is not only important for gaining insight into fundamental network problems such as network centrality measure, spectral partitioning, development of approximation algorithms, but also is crucial for understanding many real-world phenomena such as disease spreading, criticality in brain network dynamics. For a network, an eigenvector is said to be localized when most of its components take value near to zero, with a few components taking very high values. In this article, we devise a methodology to construct a principal eigenvector (PEV) localized network from a given input network. The methodology relies on adding a small component having a wheel graph to the given input network.
By extensive numerical simulation and an analytical formulation based on the largest eigenvalue of the input network, we compute the size of the wheel graph required to localize the PEV of the combined network. Using the susceptible-infected-susceptible model, we demonstrate the success of this method for various models and real-world networks consider as input networks. We show that on such PEV localized networks, the disease gets localized within a small region of the network structure before the outbreaks. The study is relevant in controlling spreading processes on complex systems represented by networks. 
}

\begin{document}

\maketitle
\nolinenumbers
\section{Introduction}
Cities are the heart of a country's economic growth, and it is estimated that by 2030, more than 60\% of the world's population will live in cities \cite{disease_spreading_2004,pandemic_influenza_2005}. We can assume cities as complex systems which evolve with time and which can be modeled using network framework. Cities embedded through basic components like infrastructure, transportation, social interaction, and energy services where individual components itself are complex systems, create several challenges as well as opportunities for network science to investigate diverse phenomena such as human mobility, stability of power-grids and spreading dynamics \cite{human_mobility_2008,power_grid_2013,disease_spreading_2004}. This article considers an epidemic spreading model, which is the basis for a large class of diffusion processes like information spreading, opinion dynamics, rumor spreading, emotional spreading in emergencies, a spread of cultural norms, diffusion of viruses, knowledge, and innovations running in the urban areas \cite{rev_dynamical_process_2012,optimal_deployment_spreading_2017}. These dynamical processes have an impact on how cities evolve or behave as a larger complex system. The current explosive trend in urbanization raises important concerns related to the public health issue. For example: how can we develop strategies to slow an initial spread of an epidemic, providing sufficient time for developing and successfully employs a vaccine \cite{disease_spreading_2004,pandemic_influenza_2005,urbanization_hantavirus_2018,urbanization_influenza_2018}. Implementation of optimized strategies have been shown to control epidemic spreads saving the loss of life and nature \cite{sudden_oak_death_2016}. 

A cornerstone feature of epidemic processes (e.g., SIS model) is the presence of the so-called epidemic threshold \cite{rev_dynamical_process_2012}. Below the epidemic threshold, the disease lies in the endemic state, and above the threshold, the disease spreads across the population. It is well known that the threshold is inversely proportional to the largest eigenvalue of the adjacency matrix of the underlying network \cite{rev_dynamical_process_2012} and a spreading process slows down near the vicinity of the threshold if the corresponding eigenvector, referred as PEV, is localized \cite{Goltsev_prl2012,metastable_loc_2016,mieghem_loc_epidemic_2018,pastor_loc_epidemic_2018}. 
Furthermore, eigenvector localization properties of complex networks are important for gaining 
insight into fundamental network problems such as networks centrality measure, spectral partitioning, development of approximation algorithms \cite{loc_centrality_2014,localization_in_mat_2016} etc., as well as dynamical phenomenon of a network such as a criticality in brain network
dynamics \cite{brain_networks2013}. Localization of PEV of a network refers to a state when a few components of the vector take very high values while the rest of the components take small values. Localizing the PEV can lead to localization of the spreading dynamics \cite{Goltsev_prl2012} on the corresponding network, as it indicates that few nodes impart huge contributions for a linear-dynamical process with a negligible contribution from rest of the nodes. 

In the current study, we present a method to localize the PEV of a given network having delocalized PEV by adding a small sub-graph to it.
Using the largest eigenvalue of the given input network, we analytically calculate the sub-graph size required for PEV localization of the combined network structure. We demonstrate that analytically calculated size is in good agreement with that of the numerical results. Finally, by running the SIS model on such PEV localized network, we show that before the disease becomes pandemic, it stays localized on the few nodes of those networks demonstrating the effectiveness of our method. We perform the experiment with the model networks as well as networks generated from various real-world data. Note that we consider connected networks, hence in the steady-state disease becomes pandemic. However, the interesting observation is that in PEV localized networks, epidemic outbreaks need more time to spread over the networks.

The article is organized as follows: section 2 contains notations, definitions, and background of the eigenvector localization. Section 3 illustrates the procedure for localizing the PEV, followed by demonstrating the success of the methodology by considering the SIS model on PEV localized networks. Finally, section 4 summarizes the study and discusses open problems for further investigation.
\begin{figure}[t]
\begin{center}
\includegraphics[width=3.4in, height=1.1in]{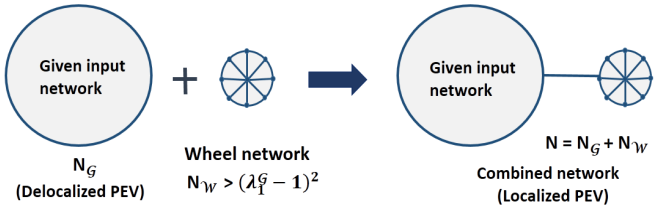}
\caption{Schematic diagram representing the construction of PEV localized network for a given network by combining a wheel network to it. For a given input network of size $N_\mathcal{G}$ with largest eigenvalue ($\lambda_1^{\mathcal{G}}$), we add a wheel graph of size $N_\mathcal{W}>(\lambda_1-1)^2$ which yields a combined graph having localized PEV.}
\label{schematic_diagram}
\end{center}
\end{figure}

\section{Model and Techniques}
We represent a graph (network) as $\mathcal{G} =\{V,E\}$ where $V=\{v_1, v_2,\ldots,v_n\}$ represents the set of vertices and $E=\{e_1, e_2,\ldots,e_{m}\}$ represents the set of edges. Here, we consider $|V|=n$, and $|E|=m$. We denote the adjacency matrix corresponding to $\mathcal{G}$  as ${\bf A} \in \Re^{n \times n}$ which is defined as $(a)_{ij} = 1$, if  $v_i$ and $v_j$ are connected by an edge and $0$ otherwise. The degree of a node and the average degree can be represented as $k_{v_i}\equiv k_{i}=\sum_{j=1}^n a_{ij}$, and $\langle k \rangle = \frac{1}{n}\sum_{i=1}^n k_{i}$, respectively. Further, a sub-graph $\mathcal{G}^{'}$ of a graph $\mathcal{G}$ is a graph $\mathcal{G}^{'}$ whose vertex set and edge set are subsets of those of $\mathcal{G}$ \cite{miegham_book2011}. We restrict our investigation for connected, undirected, unweighted and simple (without multiple edges and self-loops) networks. Therefore, ${\bf A}$ is a real symmetric matrix and contain real eigenvalues $\{\lambda_1, \lambda_2, \ldots, \lambda_n\}$, referred as \textit{spectrum} of $\mathcal{G}$. Further, without loss of generality we can order the eigenvalues of ${\bf A}$ as $\lambda_1 > \lambda_2 \geq \cdots \geq \lambda_n$ and corresponding orthonormal eigenvectors as ${\vect{x}_1, \vect{x}_2, \cdots, \vect{x}_n}$. Additionally, we know from the Perron-Frobenius theorem \cite{miegham_book2011} that for a connected network all the entries in PEV of ${\bf A}$ is positive.

We use the inverse participation ratio (IPR) to measure the extent of PEV localization \cite{Goltsev_prl2012}. We calculate the IPR of PEV ($\bm{x}_1=((x_1)_1,(x_1)_2,\ldots,(x_1)_n)$) \cite{Goltsev_prl2012,evec_localization_2017} as follows; 
\begin{equation} \label{eq_IPR}
Y_{\bm{x}_{1}}=\sum_{i=1}^n (x_{1})_{i}^4 
\end{equation}
where $(x_{1})_{i}$ is the $ith$ component of ${\bm{x}_1}$. A complete delocalized eigenvector with components $(\frac{1}{\sqrt{n}},\frac{1}{\sqrt{n}},\ldots,\frac{1}{\sqrt{n}})$ has $Y_{{\bm{x}_1}}=\frac{1}{n}$, whereas a complete localized eigenvector with components $(1,0,\ldots,0)$ yields an IPR value equal to $Y_{\bm{x}_1} = 1$. A network is said to be regular if each node has the same degree \cite{miegham_book2011} and for any regular graph (Theorem 6 \cite{miegham_book2011}), we get ${\bm{x}_1}=(\frac{1}{\sqrt{n}},\frac{1}{\sqrt{n}},\ldots,\frac{1}{\sqrt{n}})$. Hence, $Y_{\bm{x}_1}=\frac{1}{n}$, corresponds to the complete delocalized PEV. Next, if we consider a disconnected graph where each node is isolated from each other and contains only a self-loop, the corresponding adjacency matrix will be a $n \times n$ identity matrix. For this disconnected network we can choose ${\bm{x}_1}=(1,0,\ldots,0)$ yielding $Y_{\bm{x}_1} = 1$. However, for a connected, undirected and unweighted network, IPR value lies between $ 1/n \leq Y_{\bm{x}_{1}}<1$ for $n \geq 2$. Therefore, finding a network architecture for a given $n$ with delocalized PEV is easier than  finding a connected network structure with a localized PEV \cite{evec_localization_2017}. 

\begin{figure}[t]
\begin{center}
\includegraphics[width=3.4in, height=2.7in]{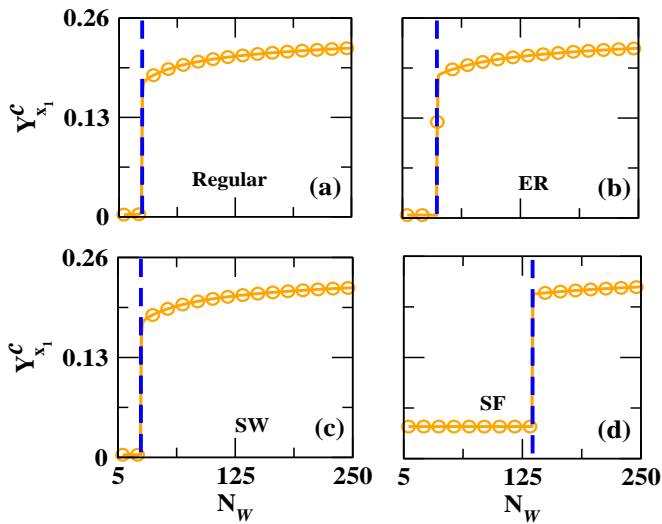}
\caption{Network structure having localized PEV. For a given network structure ($\mathcal{G}$ of size $N_\mathcal{G}$), we join with a smaller size wheel sub-graph ($\mathcal{W}$). As we vary the wheel sub-graph size ($N_\mathcal{W}$), there is an abrupt changes in the IPR value ($Y_{\bm{x}_1}^{c}$) of the combined network ($\mathcal{C}=\mathcal{G}+\mathcal{W}$, $N=N_\mathcal{G}+N_\mathcal{W}$) leads to delocalized to localized PEV state.
We choose $\mathcal{G}$ as (a) regular (b) Erd\"os-R\'enyi (ER) random network (c) small-world (SW) network and (d) scalefree network (SF) for the constructuion of PEV localized networks. We consider $N_\mathcal{G}=500$ and $\langle k \rangle=6$.} 
\label{wheel_size_vs_IPR_final}
\end{center}
\end{figure}

\begin{figure}[t]
\begin{center}
\includegraphics[width=3.2in, height=2.7in]{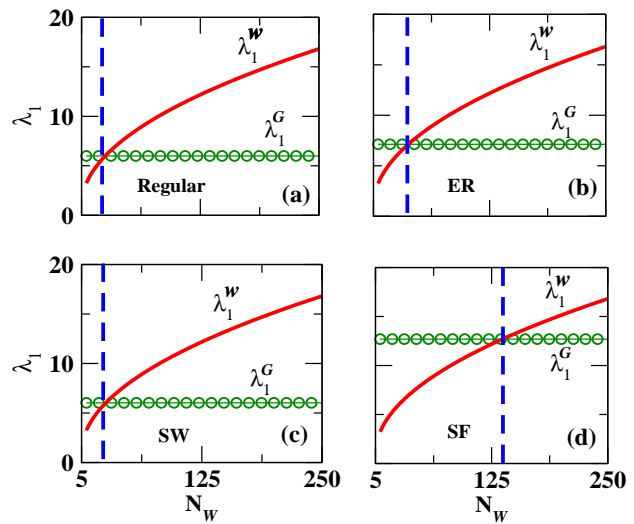}
\caption{Largest eigenvalue of the wheel sub-graph and the given input graph as a function of wheel sub-graph size. As size of
the wheel graph component increases, there is a cross over of the former over the later at a particular value  of $N_\mathcal{W}$ marked by the vertical line. (a)  regular (b) ER (c) SW and (d) SF networks. The input network is fixed, hence $\lambda_1^{\mathcal{G}}$ is also fixed. However, an increase in $N_\mathcal{W}$ leads to an increase in $\lambda_1^{\mathcal{W}}$. Network parameters are same as in Fig. \ref{wheel_size_vs_IPR_final}.}
\label{eigenval}
\end{center}
\end{figure}
\section{Results and Discussions}
We present a scheme for PEV localization of a given network. The scheme show that for a given network structure, adding a small size wheel sub-graph to the given graph leads to PEV localization of the combined graph \cite{evec_localization_2017,evec_localization_2018}. 
First, we demonstrate the success of the scheme by localizing PEV of various given model networks followed by considering few real-world networks. Additionally, by running the SIS model on the combined graph, we show that the disease stays localized within a small region before the epidemic outbreaks.

\subsection{Construction of PEV localized network from a given input network through adding of wheel graph}\label{pev_loc}
\begin{figure}[t]
\begin{center}
\includegraphics[width=3.3in, height=1.4in]{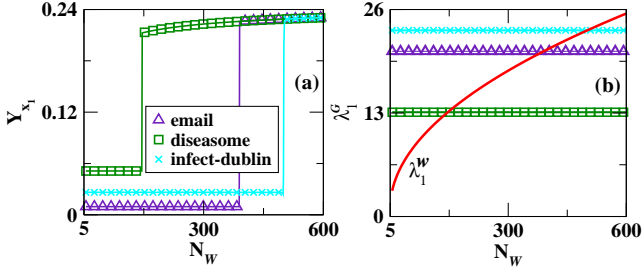}
\caption{Network structure having localized PEV. (a) For a given real-world network structure ($\mathcal{G}$ of size $N_\mathcal{G}$), we join with a smaller size wheel sub-graph ($\mathcal{W}$). As we vary the wheel sub-graph size ($N_\mathcal{W}$), there is an abrupt changes leading to localized PEV state for the combined graph ($\mathcal{C}=\mathcal{W}+\mathcal{G}$). (b) Horizontal lines are the largest eigenvalue ($\lambda_1^{\mathcal{G}}$) of the given network and the line passes through the horizontal lines is the largest eigenvalue of the wheel sub-graph ($\lambda_1^{\mathcal{W}}$). Here, we consider a email network ($N_\mathcal{G}=1133$, $\langle k \rangle=9.62$), diseasome network ($N_\mathcal{G}=516$, $\langle k \rangle=4$) and infect-dublin proximity network ($N_\mathcal{G}=410$, $\langle k \rangle=13$) data \cite{matrix_collections_2011,diseasome_network_2007,network_repository_2015}.}
\label{wheel_size_vs_IPR_real}
\end{center}
\end{figure}
A wheel graph is denoted as $\mathcal{W}=\{V_\mathcal{W}, E_\mathcal{W}\}$ and formed by connecting one node to all the nodes of a cycle graph of size $n-1$. We denote $|V_\mathcal{W}|=N_\mathcal{W}$ is the number of nodes and $|E_\mathcal{W}|=2(N_\mathcal{W}-1)$ is the number of edges and minimum size wheel graph contains $|V_\mathcal{W}|=4$ and $|E_\mathcal{W}|=6$. We know that for a regular network, PEV is delocalized. We construct a PEV localized network by combining a wheel sub-graph of size $N_\mathcal{W}=5$ with a regular network of size $N_{\mathcal{R}}$ and average degree, $\langle k \rangle=r$. 

As the size of the wheel sub-graph increases (Fig. \ref{wheel_size_vs_IPR_final}(a)), there exists an abrupt change in the IPR value of PEV of the combined network, and the IPR takes a high value indicating localization of PEV. The PEV stays localized with a further increase in the size of the wheel sub-graph (Fig. \ref{wheel_size_vs_IPR_final}(a)). Similarly, instead of considering regular network, if we consider Erd\"os-R\'enyi (ER) random, small-world (SW), or scale-free (SF) model networks \cite{newman2010}, we observe the same PEV localization phenomenon in the combined network structure (Fig. \ref{wheel_size_vs_IPR_final}(b-d)). The ER random network is generated with an edge probability $p=\langle k \rangle/n$, where $\langle k \rangle$ is the average degree of the network. To construct SW networks, we use the Watts-Strogatz model, and the SF network is constructed using the Barabasi-Albert preferential attachment method \cite{newman2010}. 

One can observe that for a particular size of the added wheel sub-graph, PEV of the combined network becomes localized. The size of the wheel sub-graph required to make the PEV of the combined graph localized can be different for different input graphs (Fig. \ref{wheel_size_vs_IPR_final}). 
In fact, it turns out that there exists a relationship between the wheel sub-graph size and the input graph properties to make localization of PEV for the combined graph. Taking a clue from the existence of a relationship between the PEV localization of networks and the largest eigenvalue of the  corresponding adjacency matrices \cite{evec_localization_2017}, we track the largest eigenvalue ($\lambda_1^{\mathcal{W}}$ and $\lambda_1^{\mathcal{G}}$) of the isolated wheel sub-graph and input network with $N_{\mathcal{W}}$ during the numerical simulations (Fig. \ref{eigenval}).
We notice that size of the wheel sub-graph for which $\lambda_1^{\mathcal{W}}$ crosses $\lambda_1^{\mathcal{R}}$ (Fig. \ref{eigenval}(a)), IPR of the PEV of the combined network structure jumps to a large value indicating the localized state and remains constant with the increase in $N_{\mathcal{W}}$ confirms localization of PEV (dotted vertical line in Figs. \ref{wheel_size_vs_IPR_final}(a) and \ref{eigenval}(a)). Thus, as soon as $\lambda_1^{\mathcal{W}} > \lambda_1^{\mathcal{R}}$ is satisfied, PEV of the combined network becomes localized.  
In the following, we analytically calculate the size of the wheel sub-graph for which the combined network gets localized. 
We know that for the wheel sub-graph, largest eigenvalue is $\lambda_1^{\mathcal{W}} = 1+\sqrt{N_\mathcal{W}}$ \cite{eigval_wheel_graph} which depends on the size of the wheel sub-graph. Hence, by substituting $\lambda_1^{\mathcal{W}} = 1+\sqrt{N_\mathcal{W}}$ in $\lambda_1^{\mathcal{W}} > \lambda_1^{\mathcal{R}}$, we get the minimal size of the wheel sub-graph 
\begin{equation}
N_\mathcal{W}> (\lambda_1^{\mathcal{R}}-1)^2, 
\end{equation}
which should hold true to make PEV of the combined network localized. We know that for a regular network, the largest eigenvalues is $\lambda_1^{\mathcal{R}} = r$, which is equal to the average degree ($\langle k \rangle=r$) of the network. Hence, size of the wheel sub-graph is $N_\mathcal{W}> (r-1)^2$ which is required to make PEV of the combined network localized. This relation is also be verified from Figs. \ref{wheel_size_vs_IPR_final}(a) and \ref{eigenval}(a). In the similar manner, we can calculate the size of the wheel sub-graph for which PEV of the combined graph localized for different input graphs such as ER, SW and SF model networks (Fig. \ref{eigenval}(b-d)).

Finally, we present the results for a few real-world networks namely, email, diseasome and infect Dublin \cite{matrix_collections_2011,diseasome_network_2007,network_repository_2015}). We perform the same experiment by connecting a wheel sub-graph of small sizes ($N_{\mathcal{W}}=5$), and we observe the transition of localized PEV state for the combined network upon increasing the size of the wheel sub-graph (Fig. \ref{wheel_size_vs_IPR_real}(a)). Fig. \ref{wheel_size_vs_IPR_real}(b) illustrates that when the largest eigenvalue of the wheel sub-graph crosses the largest eigenvalue of the given input real-world network, there exists an abrupt change in the IPR value of PEV bringing it into a localized state from a delocalized state. Again, the minimal size of the wheel sub-graph required to localize the PEV of the combined network is in good agreement to that predicted by the relation $N_\mathcal{W}> (\lambda_1^{\mathcal{G}}-1)^2$  where $\lambda_1^{\mathcal{G}}$ is the largest eigenvalue of the given input network (Fig. \ref{schematic_diagram}). Note that we always connect the peripheral node of the wheel graph to the smaller degree node of the given input network, as depicted in Fig. \ref{schematic_diagram}.

\begin{figure}[t]
\centering
\includegraphics[width=3.4in, height=2.8in]{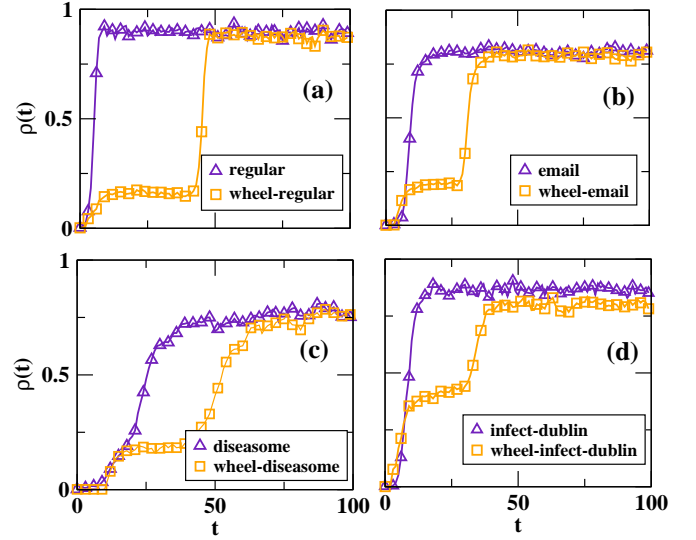}
\caption{Spreading process of SIS model on the (a) regular and combined wheel-regular, (b) email and combined wheel-email, (c) diseasome, and combined wheel-diseasome, and (d) infect-Dublin and combined wheel-infect-Dublin proximity networks are portrayed. Here, $\rho(t)$ is the fraction of nodes infected at time $t$ and size of the wheel sub-graph, $N_{\mathcal{G}}>(\lambda_1^{\mathcal{G}}-1)^2$ where $\lambda_1^{\mathcal{G}}$ is the largest eigenvalue of the given network structure. For regular, email, diseasome, and infect-Dublin networks most of the nodes are infected for smaller $t$ value, however, for combined networks having localized PEV, the disease infects small fraction of nodes and stay localized within few nodes and later becomes pandemic as $t$ increases.}
\label{spreading_real_world}
\end{figure}

\subsection{SIS model on the constructed PEV localized network}
To illustrate the success of our method for localization of epidemic spread, we use the standard SIS model \cite{Goltsev_prl2012}. In the SIS model, each susceptible vertex becomes infected with an infection rate $\gamma$, and infected vertices become susceptible to the unit rate. With a probability $\rho_{i}(t)$, a vertex $i$ gets infected by its neighbours at time $t$, and is described by the following evolution equation \cite{Goltsev_prl2012}
\begin{equation}
 \frac{d\rho_i(t)}{dt} = -\rho_i(t) + \gamma [1-\rho_i(t)]\sum_{j=1}^n a_{ij}\rho_j(t)
\end{equation}
The fraction of nodes infected at a given time $t$ is given by $\rho(t) = \sum_{i=1}^n \rho_{i}(t)/n$. We know that as the infection rate $\gamma$ crosses the epidemic threshold, i.e., $\gamma>\gamma_c$, the disease will spread over the entire network. It is known that the epidemic threshold is inversely proportional to the largest eigenvalue of the corresponding adjacency matrix i.e., $\gamma_c=\frac{1}{\lambda_1}$ \cite{rev_dynamical_process_2012}. However, if PEV of the adjacency matrix is localized, in the vicinity of the epidemic threshold $\gamma_c + \epsilon, \; \epsilon>0$ the disease infects a small number of vertices, and the spreading process slows down \cite{Goltsev_prl2012}. Therefore, it requires a large time for the disease to spread over the entire network. We perform the SIS model on a network where initially all the nodes are susceptible except one single node, which is chosen at random and made infected. With $\gamma$ ($>\gamma_c$) rate, the infected node infects the adjacent susceptible nodes, and with a unit rate, the infected nodes become susceptible again. This process is repeated for a large time ($t$) until the steady-state is reached. 

We run the SIS dynamics on the model as well as on real-world networks and find that the epidemic spread becomes pandemic within a very short period, whereas upon suitable addition of a wheel subgraphs to the network disease becomes localized (Fig. \ref{spreading_real_world}). For example, Fig. \ref{spreading_real_world}(a) manifests that for the SIS model on a regular network, the disease infects a large number of vertices and becomes pandemic within a very short period. Whereas, for the wheel-regular network structure, the disease stays localized within a few nodes before the outbreaks (Fig. \ref{spreading_real_world}(a)). Fig. \ref{spreading_real_world}(b-d) present results for real-world networks and show that one can localize the epidemic spread by combining a suitable size wheel sub-graph structure. By connecting the real-world network structure with a minimal size wheel sub-graph as discussed in the previous section, we convert the input network structure into a PEV localized network. All the codes developed in this paper and data are available at GitHub repository \cite{codes_data_wheel_graph_strategy}. 

\section{Conclusions}
To conclude, the study provides a methodology to convert a given PEV delocalized input graph to PEV localized graph by the addition of a small wheel sub-graph. First, we numerically examine the success of the methodology by investigating the spread of epidemics via the SIS model on such PEV localized networks. The investigation demonstrates that before the outbreak, the disease stays localized within a few nodes. Second, analytically we derive a relationship between the size of the wheel graph required for PEV localization of the combined graph and the largest eigenvalue of the given input graph. 

There may exist other methods to convert a given delocalized network to a localized network. For example, Ref. \cite{evec_localization_2017} provides a method to get PEV localized network, which is based on the rewiring of the given network edges. However, it may not always be feasible to perform the rewiring of edges of real-world networks. The approach presented here shows that one can convert a given real-world network into PEV localized graph by adding a small subgraph. Since the size of the added wheel sub-graph depends on the largest eigenvalue of the input network, and most of the real-world networks are sparse, thereby yielding a rather small value of the largest eigenvalue. Therefore, conversion to the PEV localized graph needs a small size wheel graph, making our methodology relevant for real-world applications. 

Here, we have considered the SIS model, but the formulation is useful in understanding diffusion processes for other related models such as susceptible-infected-recovered (SIR) epidemic model, rumor propagation, simple models mimicking the routing of information packets in technological systems, reaction-diffusion processes, etc. \cite{rev_dynamical_process_2012}. Further, this work is restricted only to the single-layer network model. Many real-world systems can be better modeled using a multilayer network framework. Hence, to understand diffusion processes on multilayer networks and make them PEV localized is an interesting problem which requires further investigation \cite{multilayer_evec_loc_2018,multilayer_disease_loc_2017}. There exists other ways to construct PEV localized network structure mainly focused on hub node and interested readers are refers to study and investigate SIS model on them \cite{loc_centrality_2014,evec_centrality_2019,loc_bipartite_2017,disease_localization_2018}. However, in this article, we furnish a method of constructing PEV localized networks through a wheel graph instead of a hub node taking a clue from \cite{evec_localization_2017} and provides a platform to understand the localization of disease spread in networks.

\acknowledgments
SJ acknowledges CSIR (25(0293)/18/EMR-II), and DAE (37(3)/14/11/2018-BRNS/37131) Govt. of India grants for financial support. PP acknowledges CSIR, Govt. of India grant (09/1022(0070)/2019-EMR-I) for Senior Research Fellowship.

\section{Appendix: Wheel-Random-Regular model}\label{appendix}
We know that connecting a wheel network ($\mathcal{W}$) with an edge to a random regular network ($\mathcal{R}$) as in Fig. \ref{schematic_diagram}, from Ref. \cite{evec_centrality_loc_2020}(Appendix B, Eq. (B.7)) we get
\begin{equation}\label{relation1}
\lambda_1^\mathcal{W} - \lambda_1^\mathcal{R} = \frac{\bm{v}_{1}^{\mathcal{R}^{T}} \mathcal{P}^{T}\bm{x}_{1}^1}{\bm{v}_{1}^{\mathcal{R}^{T}}\bm{x}_{1}^2}  - \frac{\bm{v}_{1}^{\mathcal{W}^{T}} \mathcal{P}\bm{x}_{1}^2}{\bm{v}_{1}^{\mathcal{W}^{T}}\bm{x}_{1}^1}  
\end{equation}
where $\bm{v}_{1}^{\mathcal{W}}=(\frac{1}{\beta},\frac{\alpha}{\beta},\ldots,\frac{\alpha}{\beta})^{T}$ such that $\alpha=\frac{\sqrt{n_{1}}+1}{n_{1}-1}$, $\beta=\sqrt{1+\frac{(\sqrt{n_{1}}+1)^2}{n_{1}-1}}$ \cite{eigval_wheel_graph} and $\bm{v}_{1}^{\mathcal{R}}=(\frac{1}{\sqrt{n_2}},\frac{1}{\sqrt{n_2}},\ldots,\frac{1}{\sqrt{n_2}})^{T}$ are the PEV of $\mathcal{W}$ and $\mathcal{R}$, respectively. Moreover, we denote $n_1 \equiv N_{\mathcal{W}}$ and $n_2 \equiv N_{\mathcal{R}}$. Here, $\vect{x}_1=(\vect{x}_1^1|\vect{x}_1^2)^{T}$ is the PEV of the combined network such that $\vect{x}_1^1=((x_1^1)_1,(x_1^1)_2,\ldots,(x_1^1)_{n_1})^{T}$ and $\vect{x}_1^2=((x_1^2)_1,(x_1^2)_2,\ldots,(x_1^2)_{n_2})^{T}$. For $n_1>(r-1)^2$  or $\lambda_1^\mathcal{W} > \lambda_1^\mathcal{R}$
\begin{equation}\nonumber
\begin{split}
\frac{\bm{v}_{1}^{\mathcal{W}^{T}} \bm{x}_{1}^1}{\bm{v}_{1}^{\mathcal{R}^{T}} \bm{x}_{1}^2} &> \frac{\bm{v}_{1}^{\mathcal{W}^{T}} \mathcal{P}\bm{x}_{1}^2}{\bm{v}_{1}^{\mathcal{R}^{T}} \mathcal{P}^{T}\bm{x}_{1}^1} \\
\end{split} 
\end{equation}
\begin{equation}\nonumber
\begin{split}
\frac{(x_{1}^{1})_{n_1}[(\sqrt{n_1}-2)(x_{1}^{1})_1  + \sum_{i=1}^{n_1} (x_{1}^{1})_i]}{(x_{1}^{2})_1 \sum_{i=1}^{n_2} (x_{1}^{2})_i}  &> 1 \\
\end{split} 
\end{equation}
From the above relation one can say that holding the relation $\lambda_{1}^{\mathcal{W}} > \lambda_{1}^{\mathcal{R}}$, PEV of the combined network for which maximum contribution comes from the wheel graph part. Now, instead of wheel graph if we consider a star network then for $\lambda_1^\mathcal{S} > \lambda_1^\mathcal{R}$, from Eq. (\ref{relation1}) we get
\begin{equation}\nonumber
\frac{\bm{v}_{1}^{\mathcal{S}^{T}} \bm{x}_{1}^1}{\bm{v}_{1}^{\mathcal{R}^{T}} \bm{x}_{1}^2} > \frac{\bm{v}_{1}^{\mathcal{S}^{T}} \mathcal{P}\bm{x}_{1}^2}{\bm{v}_{1}^{\mathcal{R}^{T}} \mathcal{P}^{T}\bm{x}_{1}^1} 
\end{equation}
\begin{equation}\nonumber
\frac{\sqrt{(n_1-1)}(x_{1}^{1})_1 (x_{1}^{1})_{n_1} + (x_{1}^{1})_{n_1} \sum_{i=2}^{n_1} (x_{1}^{1})_i}{(x_1^2)_1 \sum_{i=1}^{n_2} (x_{1}^{2})_i}>1 
\end{equation}
where $\bm{v}_{1}^{\mathcal{S}}=\biggl(\frac{1}{\sqrt{2}},\frac{1}{\sqrt{2(N_{\mathcal{S}}-1)}},\ldots,\frac{1}{\sqrt{2(N_{\mathcal{S}}-1)}}\biggr)$ is the PEV of star network \cite{miegham_book2011}. Now, we can extend the results in Eq. (\ref{relation1}) for other graphs instead of regular graph. For  $N_\mathcal{W} > (\lambda_1^\mathcal{G}-1)^2$ or $\lambda_1^\mathcal{W} > \lambda_1^\mathcal{G}$ we have
\begin{equation}\nonumber
\frac{\bm{v}_{1}^{\mathcal{W}^{T}} \bm{x}_{1}^1}{\bm{v}_{1}^{\mathcal{G}^{T}} \bm{x}_{1}^2} > \frac{\bm{v}_{1}^{\mathcal{W}^{T}} \mathcal{P}\bm{x}_{1}^2}{\bm{v}_{1}^{\mathcal{G}^{T}} \mathcal{P}^{T}\bm{x}_{1}^1} 
\end{equation}
\begin{equation}\nonumber
\frac{v_1(x_1^1)_1(x_1^1)_{n_1} +\alpha v_1(x_1^1)_{n_1}\sum_{i=2}^{n_1} (x_1^1)_i}{\alpha(x_1^2)_1\sum_{i=1}^{n_2} v_i(x_1^2)_i} > 1
\end{equation}
where $\bm{v}_{1}^{\mathcal{G}}=(v_1,v_2,\ldots,v_{N_{\mathcal{G}}})^{T}$ is the PEV of the given input network. Hence, if we know the PEV of the given graph, one can easily check the contribution of the individual graph component to the IPR value of the combined networks.

\end{document}